\documentclass[twocolumn,superscriptaddress, prl,showpacs,longbibliography]{revtex4-1}
\usepackage{amsfonts}
\usepackage{amssymb}
\usepackage{amsmath}
\usepackage{amsthm}
\usepackage{dsfont}
\usepackage{graphicx}
\usepackage{stackrel}
\usepackage{color}
\usepackage{bm}
\usepackage{ulem}

\usepackage{hyperref}
\hypersetup{
    colorlinks,
    citecolor=red,
    filecolor=blue,
    linkcolor=blue,
    urlcolor=blue
}

\newcommand{\ba}{\begin{align}}
\newcommand{\ea}{\end{align}}

\newcommand{\tc}{\tilde{c}}

\newcommand{\eps}{\epsilon}

\newcommand{\beg}{\begin{equation}}
\newcommand{\en}{\end{equation}}

\newcommand{\eref}[1]{Eq.~(\ref{#1})}
\newcommand{\re}[1]{(\ref{#1})}

\newcommand{\esref}[1]{Eqs.~(\ref{#1})}

\begin{document}

\title{Gaussian ensemble for quantum integrable dynamics}

\author{Hyungwon Kim}
\affiliation{Center for Materials Theory, Department of Physics and Astronomy, Rutgers University, Piscataway, NJ 08854, USA}

\author{Anatoli Polkovnikov}
\affiliation{Department of Physics , Boston University, Boston, MA 02215, USA}

\author{Emil A. Yuzbashyan}
\affiliation{Center for Materials Theory, Department of Physics and Astronomy, Rutgers University, Piscataway, NJ 08854, USA}

\begin{abstract}

We propose a Gaussian ensemble  as a description of the long-time dynamics of isolated quantum integrable systems.  Our approach extends the Generalized Gibbs Ensemble (GGE) by incorporating fluctuations of integrals of motion.  It  is asymptotically exact  in the classical limit irrespective of the system size and, under appropriate conditions, in the thermodynamic limit irrespective of the value of Planck's constant.     Moreover, it captures  quantum corrections near the classical limit,   finite size corrections     near the thermodynamic limit, and  is valid  in the presence of non-local interactions. The Gaussian ensemble  bridges the gap between classical integrable systems, where a generalized microcanonical ensemble is exact even for few degrees of freedom and  GGE, which requires  thermodynamic limit. We illustrate our results with  examples of increasing complexity.

\end{abstract}


\maketitle

The far from equilibrium dynamics of isolated  many-body systems with many nontrivial integrals of motion attracted  considerable attention as such dynamics have been recently realized in several experiments \cite{kinoshita2006,Hofferberth:2007,Gring:2012,Trotzky:2012,shimano2013,Langen:2013,Langen:2015}.
In particular, it has been conjectured that the infinite time averages of various observables for a system evolving with a time-independent Hamiltonian $\hat H$ are described by the Generalized Gibbs Ensemble (GGE) \cite{rigol1}:
\beg
\hat\rho_\mathrm{GGE}=Ce^{-\sum_i \beta_i\hat H_i},\quad h_i\equiv\langle \hat H_i \rangle_0 = \mathrm{tr}(\hat \rho_\mathrm{GGE} \hat H_i), 
\label{gge}
\en
where $\hat H_i$ is a complete (in some yet unspecified sense) set of integrals of motion for $\hat H$, the second equation  relates   $\beta_i$  to expectation values $h_i$ of the integrals in the initial state,  and $C$ is a normalization constant. A key difficulty with quantum GGE stems from the absence of an accepted well-defined  notion of quantum integrability. As a result, GGE is strictly speaking unfalsifiable. For example, it was initially shown to fail for the 1D XXZ spin chain \cite{wouters,pozsgay,goldstein}, but later studies \cite{ilievski1,ilievski2} cured this  by adding new integrals of motion in \eref{gge}.

In contrast, classical integrability is  well-defined \cite{arnold}. Moreover, the microcanonical version of GGE -- Generalized Microcanonical Ensemble (GME) is exact for   a general classical integrable Hamiltonian $H({\bm p}, {\bm q})$ \cite{arnold,emil}, 
\begin{align}
&\rho({\bm p}, {\bm q}) =  C\prod_{k=1}^n \delta(H_k({\bm p}, {\bm q}) - h_k),  \label{clGME1}\\
&\lim_{T \rightarrow \infty} \frac{1}{T}\int^T_0\!\!\! O( t) dt =  \int\! d{\bm p} d{\bm q}\, O({\bm p}, {\bm q}) \rho({\bm p}, {\bm q}), \label{clGME2}
\end{align}
where ${\bm q}=(q_1,\dots, q_n)$ and ${\bm p}=(p_1,\dots, p_n)$ are the generalized coordinates and momenta  and $H_k({\bm p}, {\bm q})$ are the integrals of motion. The time evolution of any dynamical variable (observable) $O(t)\equiv O({\bm p}(t), {\bm q}(t))$ is obtained by evolving with $H({\bm p}, {\bm q})$ starting at $t=0$ and $h_k$ is the initial value of $H_k({\bm p}, {\bm q})$. Note that unlike the microcanonical distribution for a nonintegrable Hamiltonian or GGE, \eref{clGME2} holds for any number of degrees of
freedom $n$ and arbitrary interactions, i.e. does not require thermodynamic limit. In some sense, classical integrable dynamics are more  ergodic, but in  a restricted part of the phase-space cut out by the integrals of motion.

\begin{figure}
\includegraphics[width=0.95\linewidth]{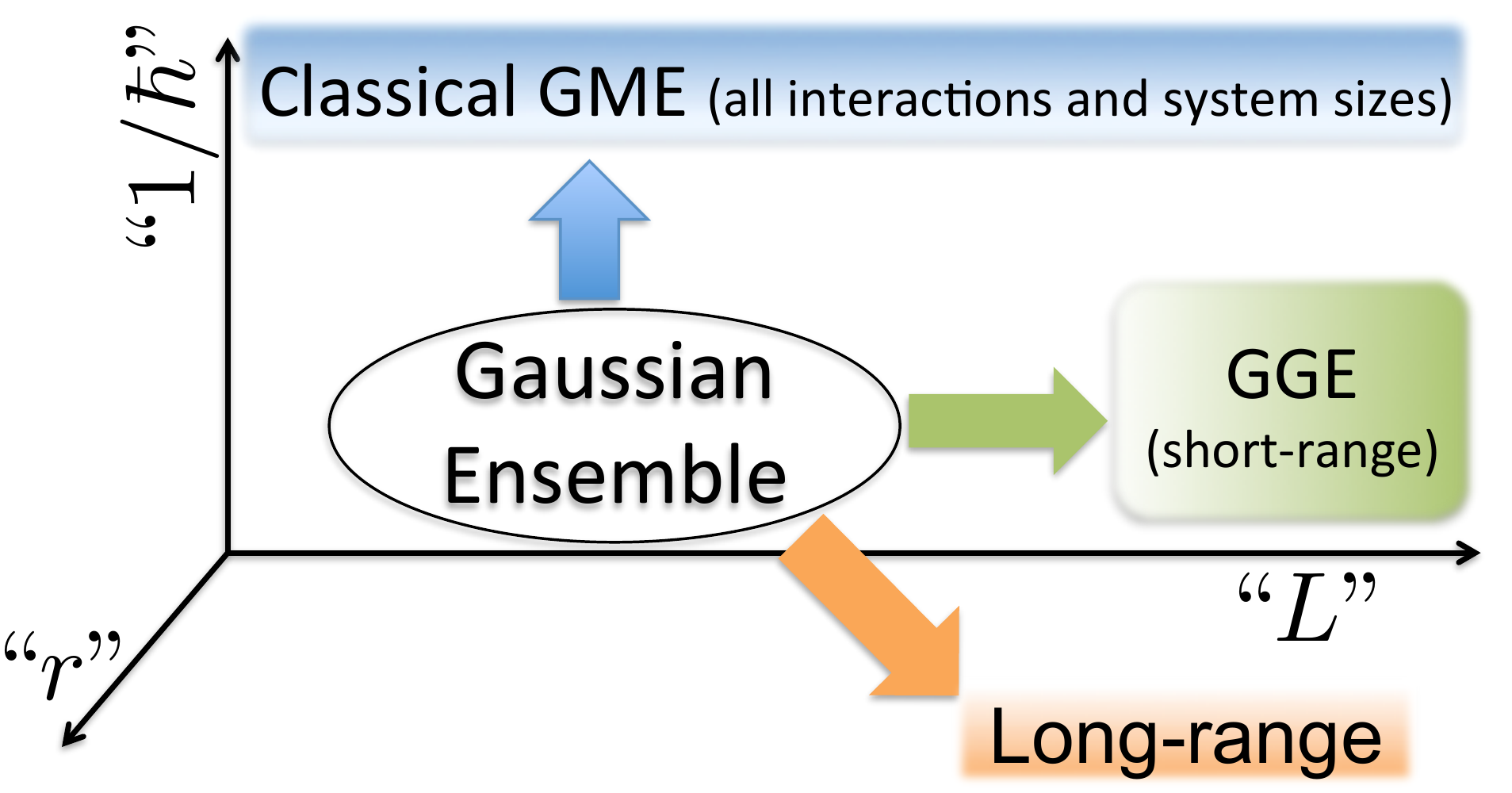}
\centering
\caption{ (color online) Schematic diagram indicating limits where the Gaussian ensemble \re{eq:GME} is exact. By design it converges to the classical GME \re{clGME1} when $\hbar\to0$. The latter is a rigorous theorem in classical mechanics valid for a general integrable system at any number of degrees of freedom. The Gaussian ensemble contains GGE as a particular case and therefore holds whenever GGE does (e.g. for short-range interactions and system size $L\to\infty$).  It is also valid for models where mean-field is exact due to large interaction range $r$. }
\label{fig:graph}
\end{figure}

 What can play the role or replace the microcanonical ensemble for a quantum integrable system with arbitrary particle number?
More preciesly, how to  quantize \esref{clGME1} and (\ref{clGME2}), i.e. what is a suitable density matrix $\hat\rho$ that turns into  \eref{clGME1} in $\hbar\to0$ limit? To what extent does it describe the quantum  dynamics and how does it compare to GGE? These are the questions we address in this paper. We propose a multivariable Gaussian in $\hat H_i$ as an appropriate  $\hat\rho$ and show that it has several remarkable features. In particular, it provides  quantum corrections to \eref{clGME1} at any particle number,    i.e. a  semiclassical approximation to the exact density matrix.   It further yields
finite size correction to GGE and is expected to work well in systems with long-range interactions, see  also Fig.~\ref{fig:graph}.

In the case of Gibbs or Generalized Gibbs distributions, one can simply replace 
$H({\bm p}, {\bm q})$ or  $H_k({\bm p}, {\bm q})$ with the corresponding operators. This  does not work for \eref{clGME1}, because the average of a product of $\hat H_i$ with so constructed density matrix is equal to the product of averages, which is not the case for a typical quantum state. Therefore, to reproduce various time averages, we need to broaden the delta-functions in \eref{clGME1}. 

It is natural to proceed by analogy with the usual quantum microcanonical ensemble and to replace the right hand side of \eref{clGME2} with an equal weight average over all eigenstates $|n\rangle$ of $\hat H_i$, $\hat H_i|n\rangle=E_i^{(n)}|n\rangle$, that have eigenvalues $E_i^{(n)}$ sufficiently close to quantum expectation values  $h_i=\langle\hat H_i\rangle_0$ of the integrals in the initial state~\cite{cassidy,he}. The problem is that $E_i^{(n)}$ are generally discrete, while $h_i$ can be anywhere in between. For example, integrals of motion for a collection of noninteracting fermions are their occupation numbers taking values 0 and 1, while their expectation values are arbitrary numbers ranging from 0 to 1. As a result, it is not always possible to find even a single eigenstate sufficiently close to the  prescribed set of $h_i$. This problem can perhaps be resolved for a certain class of models in the thermodynamic limit through coarse-graining or suitable redefinition of the integrals~\cite{cassidy,he,caux}. However, in other   integrable models there seems to be no simple, well-motivated remedy even at large particle number.  We consider one such example below -- interaction quenches in the BCS  model, where there are no eigenstates reasonably close to $h_i$ (see Fig.~\ref{fig:BCS}). In any case, since \eref{clGME2} is valid for any $n$,  we are looking for a  model-independent approach uniformly applicable at any particle number.  

Instead, we propose  broadening  $\delta$-functions in \eref{clGME1} directly. Specifically, we consider a Gaussian ensemble
\begin{align}
\hat \rho_{\mathrm{G}} =  C \exp \biggl[-\sum_{ij} ( \hat H_i - \mu_i) (\Sigma^{-1})_{ij}  (\hat H_j - \mu_j)\biggr], 
\label{eq:GME}
\end{align}
where  parameters $\mu_i$ and $\Sigma_{ij}$ are fixed by  first and second moments of the initial conditions
\begin{align}
h_i\equiv\langle \hat H_i \rangle_0 &= \mathrm{tr}(\hat \rho_{\mathrm{G}} \hat H_i), \label{eq:charge}\\
\langle \hat H_i \hat H_j \rangle_0 &= \mathrm{tr}(\hat \rho_{\mathrm{G}} \hat H_i \hat H_j ), \label{eq:correlation}
\end{align}
i.e. $\hat \rho_\mathrm{G}$  by design reproduces exact  first and second order correlation functions of conserved quantities.
Note also that when    $\hat H_i$ have unbounded and continuous spectra, $\mu_i = \langle H_i \rangle_0$ and
$\Sigma$  is the covariance matrix, $\Sigma_{ij} = \langle \hat H_i \hat H_j \rangle_0 - \langle \hat H_i 
\rangle_0 \langle \hat H_j \rangle_0$.

It turns out that this ensemble has a number of significant advantages over various alternatives. First, it converges to \eref{clGME1} and therefore is exact in the classical limit for any number of degrees of freedom. Indeed, in this limit $\langle \hat H_i \hat H_j \rangle_0 = \langle \hat H_i \rangle_0 \langle \hat H_j \rangle_0$ and spectra of $\hat H_i$ are continuous. Therefore, a product of $\delta$-functions solves \esref{eq:charge} and \re{eq:correlation} and, assuming the solution is unique, we arrive at the above statement. Moreover, we find that not only \eref{eq:GME} is exact in $\hbar\to 0$ limit, but   it also captures the leading quantum correction ($\propto \hbar$), at least in all examples we studied. Higher order corrections, however, do not necessarily agree (see  below). For this reason, we also expect the Gaussian ensemble to be exact for models  with long-ranged  interactions   in the thermodynamic limit,  such that the mean-field is  exact effectively rendering such models classical.

Second, the Gaussian ensemble contains GGE as a particular case (when the coefficients at the bilinear in $\hat H_i$ terms in the exponent in \eref{eq:GME} vanish) and therefore always works as well as or better than GGE. Furthermore, we find that averages with $\hat \rho_\mathrm{G}$ converge to exact infinite time averages faster than for GGE with increasing system size.  For example, in a 1D fermion model on $L$ sites we analyze below, the convergence is $1/L^2$ for the Gaussian ensemble and $1/L$ for GGE.

  Now suppose the only integral is the total energy and the Eigenstate Thermalization Hypothesis (ETH) holds, i.e.     expectation   values of observables in an eigenstate    are functions of the corresponding eigenenergy only \cite{Srednicki}.  Then,  the Gaussian  (unlike Gibbs or microcanonical) ensemble    reproduces not only time-averaged expectations but also fluctuations of global and local observables, since  these quantities     depend  on energy and its fluctuation only \cite{luca}.  Fluctuations in turn enter various Kubo response coefficients, fluctuation-dissipation and other thermodynamic relations.
Similarly, the multivariable Gaussian  \re{eq:GME} correctly predicts expectations and fluctuations of  observables  satisfying the generalized ETH \cite{cassidy} in  any integrable system. Lastly, \eref{eq:GME} is model-independent and  well-defined for any spectra of $\hat H_i$ with no need to choose a measure of closeness of eigenstates to the prescribed set of expectation values $h_i$.

Of course, one can consider other representations of $\delta$-functions in \eref{clGME1} and include more parameters to match higher order moments of $\{\hat H_i\}$. However, $\hat\rho_\mathrm{G}$ is sufficient for many purposes as outlined above and as we will see in the examples below. Our proposal   also does not fully resolve the difficulty with the notion of quantum integrability mentioned in the first paragraph. However, it in part bypasses this issue in models with smooth integrable classical limit, such as e.g. BCS-Gaudin model considered below. In such cases, we can rely on the well-defined concept of classical integrability and \eref{eq:GME} makes good sense due to its connection to the rigorous result \re{clGME2} for classical integrable systems.  In the absence of a sound quantum definition, this route via classical integrability is useful for making unambiguous  statements about general properties of quantum  systems believed to be integrable. For example, Ref.~\onlinecite{berry} uses this approach to derive energy level statistics in such systems.
In what follows we consider several specific examples to illustrate the above points.

{\it Convergence to classical GME.} Perhaps, the simplest quantum system with a transparent classical limit is the 1D harmonic oscillator in a coherent state. In this case, we find that the discrepancy between  exact infinite time averages and the Gaussian ensemble expectation values of observables is of order $\hbar^2$ \cite{SM}. For example,
\beg
\frac{\overline{\langle \hat{\Lambda}^k \rangle}_{\infty}}{E_0^k} =\frac{\langle \hat{\Lambda}^k \rangle_\mathrm{G}}{E_0^k}=1 + \frac{k(k-1)}{2}\frac{\hbar\omega}{E_0} + O\left(\frac{\hbar^2\omega^2}{E_0^2} \right),
\en
where $\hat\Lambda=\hbar\omega \hat n$, $\hat n$ is the number operator and $E_0$ is the classical energy.

 Next, we look at two-spin Gaudin magnets (see below for  larger number of spins) --  two interacting spins   of arbitrary magnitudes $(S_1, S_2)$ in a magnetic field,  
\begin{align}
\hat H_1 &= B \hat S_1^z + \gamma \hat {\bm S}_1\cdot\hat {\bm S}_2, \label{h1}\\
\hat H_2 &= B \hat S_2^z - \gamma \hat {\bm S}_1\cdot\hat {\bm S}_2, \quad [\hat H_1, \hat H_2]=0.
\end{align}
Let us designate $\hat H_1$ as the Hamiltonian to generate  quantum dynamics, though both Gaussian ensemble and infinite time averages are the same for a generic linear combination of $\hat H_1$ and $\hat H_2$. Classically (when spins become angular momenta variables), this is an integrable system with two degrees of freedom. There are five independent parameters in $\hat \rho_\mathrm{G}$ to be fixed by two first and three second moments of $\hat H_1$ and $\hat H_2$. 

We start the dynamics from $|\psi(0)\rangle = |{\bm \sigma}_1\rangle\otimes|{\bm \sigma}_2\rangle$, where  $|{\bm \sigma}_{i=1,2}\rangle$ is a spin coherent state  characterized by a  direction $(\theta_i, \phi_i)$   in which the projection of $ \hat {\bm S}_i$ is maximal.   We choose coherent initial states  for the ease of visualizing   dynamics in the classical limit, since  they correspond to individual points in the phase-space. However, other initial states are equally good.
For an observable, we pick $\hat S_1^z$, arbitrarily set $(\theta_1, \phi_1) = (0.5\pi, 0.5\pi)$, $(\theta_2, \phi_2) = (0.3\pi, 0)$, $\gamma = 1$, and  choose $B = \gamma S_2$, so that the effects of the magnetic field
and  interaction on the first spin are comparable.

\begin{figure}
\includegraphics[width=0.95\linewidth]{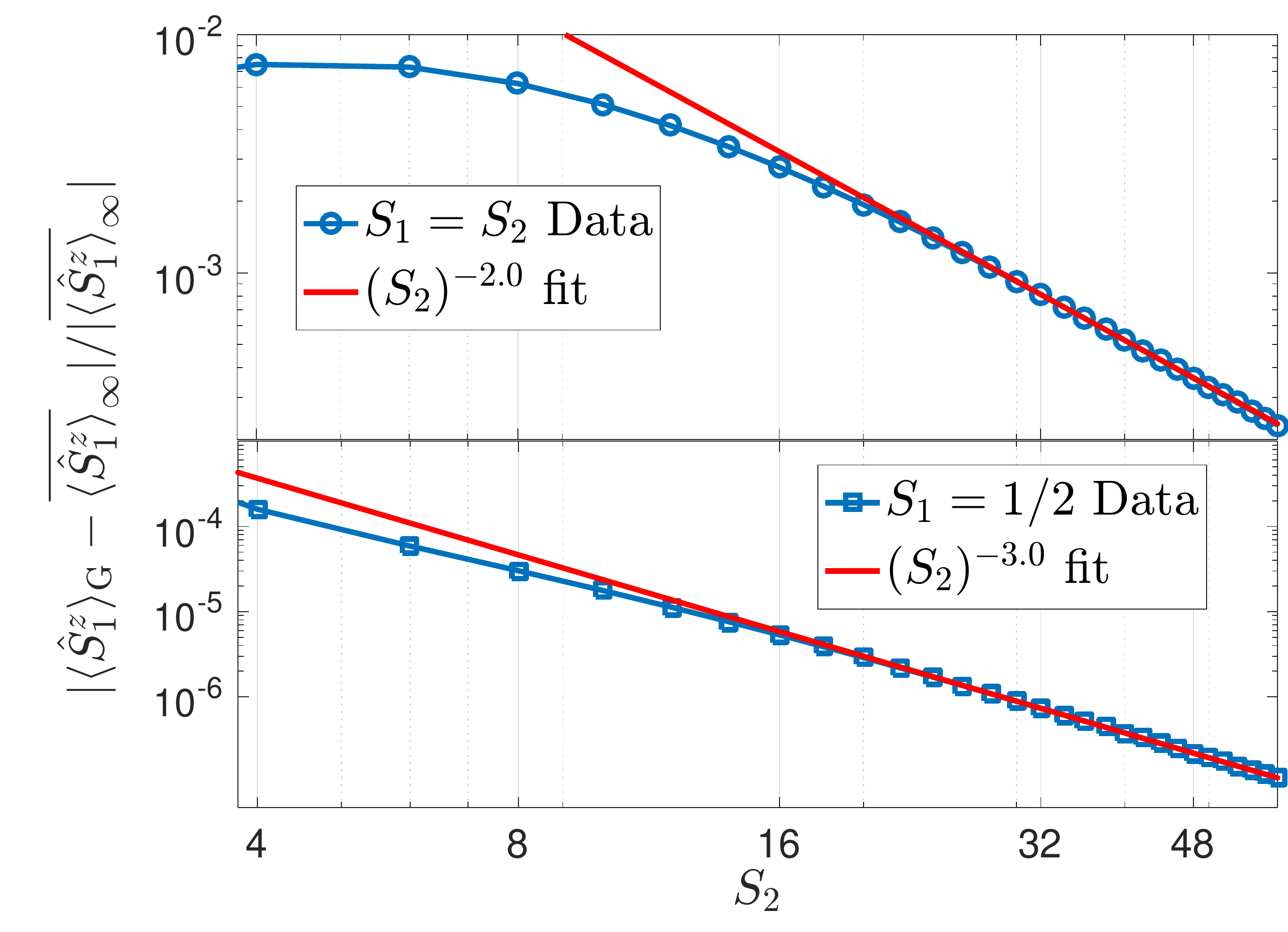}
\centering
\caption{ (color online) Normalized difference between Gaussian ensemble and infinite time averages   of $S_1^z$ for the two-spin Hamiltonian \re{h1} as a function of the  magnitude $S_2$ of the second spin.
Upper panel: $S_1 =   S_2$, lower panel: $S_1 = 1/2$. The two averages converge in the limit $\hbar\to0$, $S_2\to\infty$, $\hbar S_2=\mbox{fixed}$.   Corrections to the Gaussian ensemble  are of order $\hbar^2$ in the first case and $\hbar^3$ in the second.
 }
\label{fig:Spin}
\end{figure}

We consider two cases: $S_1=S_2$ and $S_1 = 1/2$ at 
  increasing $S_2$. The classical limit (for the second spin) is $\hbar\to0$, $S_2\to\infty$, while keeping $\hbar  S_2=\mathrm{const}$. Therefore, terms of order $1/S_2^k$ are of order $\hbar^k$. Fig.~\ref{fig:Spin} shows the difference $ \langle \hat S^z_1\rangle_\mathrm{G} -\overline{\langle \hat S^z_1\rangle}_{\infty}$ between the Gaussian ensemble and  infinite time averages as   a function of $S_2$. In the first case, the system becomes truly classical as $S_1=S_2\to\infty$ and the discrepancy (in $\hbar\langle \hat S^z_1\rangle=\mathrm{finite}$) is of order $\hbar^2$. We see that the Gaussian ensemble indeed captures the main quantum correction of order $\hbar$. Somewhat surprisingly, the agreement is even better in the second case, when the first spin stays quantum. Here at large $S_2$ the difference goes as $1/S_2^3$, i.e. the discrepancy is of order $\hbar^3$.

\begin{figure}
\includegraphics[width=0.95\linewidth]{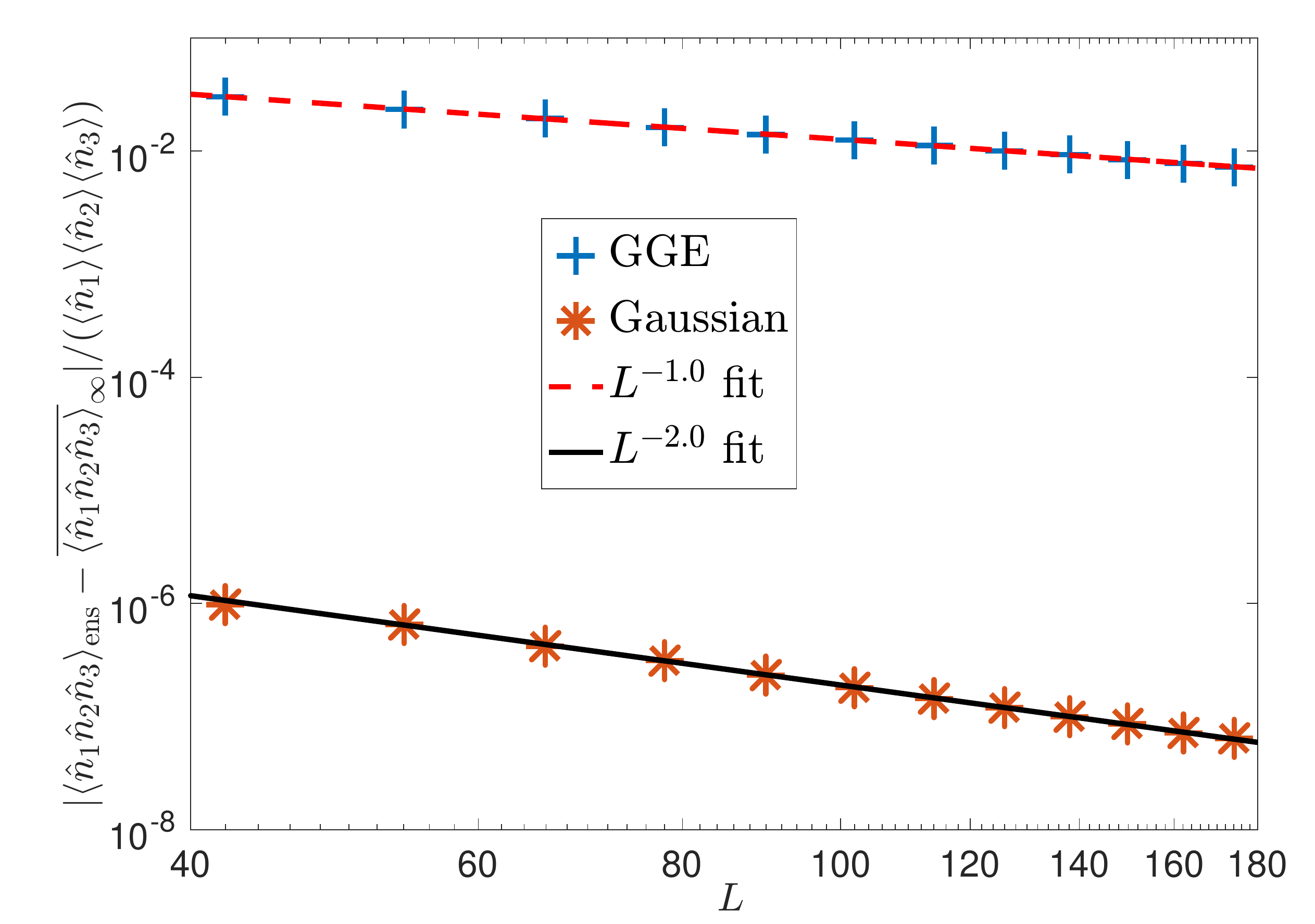}
\centering
\caption{(color online) Mismatch between ensemble (Gaussian and GGE) and infinite  time averages in a three-point   correlation function of the lattice site occupation number $\hat n_i$ for the free-fermion model \re{ff} with $Q = \pi/3$ as a function of the chain length $L$.  We normalize  by the first cumulant, which is the same for both ensembles. The time evolution is  due to a quench from $\phi, V_1, V_2=0$ to $\phi = 0.3$, $V_1 = 1.5$, $V_2 = 1.0$.  The  mismatch vanishes as $L^{-1}$ for GGE and as $L^{-2}$ for the Gaussian ensemble. The latter thus captures the leading finite size correction ($\propto L^{-1}$) to the thermodynamic limit. }
\label{fig:GGE}
\end{figure}

{\it Comparison with GGE.} To compare with GGE at increasing system size, 
 we analyze quenches in a free-fermion system, where GGE is exact in the thermodynamic limit,  
\begin{align}
\hat H = &-\sum_{j=1}^L (e^{i \phi} \hat c_{j+1}^\dag \hat c_j +e^{-i\phi} \hat c_j^\dag \hat c_{j+1}) \nonumber\\
&+ \sum_{j=1}^L\left[V_1 \cos(Q j) + V_2 \cos(2Q j)\right]\hat n_j, \label{ff}
\end{align}
where $\hat c_j$ annihilates a fermion at site $j$, $\hat  n_j = \hat  c_j^\dag \hat  c_j$, and $Q = 2\pi/M$ is a commensurate modulation. We impose a periodic boundary condition and choose $M= 6$ ($L$ is a multiple of $M$). 
We prepare the system in the ground state of $\hat H$ with $\phi = V_1 = V_2 = 0$ at half filling 
and  quench to  nonzero $\phi$, $V_1$, and $V_2$. This mixes $M$ single-particle eigenstates in the pre-quench Hamiltonian. The parameter $\phi$ breaks the time-reversal symmetry and $V_1, V_2$ break the particle-hole symmetry  thus removing all symmetry protected degeneracies in the single-particle spectrum.  
Natural integrals of motion  are mode occupation numbers of the post-quench Hamiltonian. 

Since GGE by construction captures averages of all single-particle occupation numbers and Gaussian ensemble -- those of all their linear and bilinear combinations, they exactly reproduce the time average of  single-body and two-body observables, respectively, for any $L$ \cite{SM}.  Therefore, we study a three-body correlation function $\langle \hat n_1 \hat n_2 \hat n_3\rangle$. In Fig.~\ref{fig:GGE}, we plot the difference between the infinite time and  ensemble averages  normalized by the first cumulant $\langle \hat n_1\rangle \langle \hat n_2 \rangle \langle \hat n_3\rangle$.  
We see that the GGE approaches the thermodynamic limit $L\to\infty$ as $1/L$ as expected \cite{he,gramsch,wright} while the Gaussian ensemble -- as $1/L^2$ as  anticipated above \cite{SM}.  In addition to faster approach, the Gaussian ensemble  result agrees with the time average better at any given $L$ by orders of magnitude.

 \begin{figure}
\includegraphics[width=0.95\linewidth]{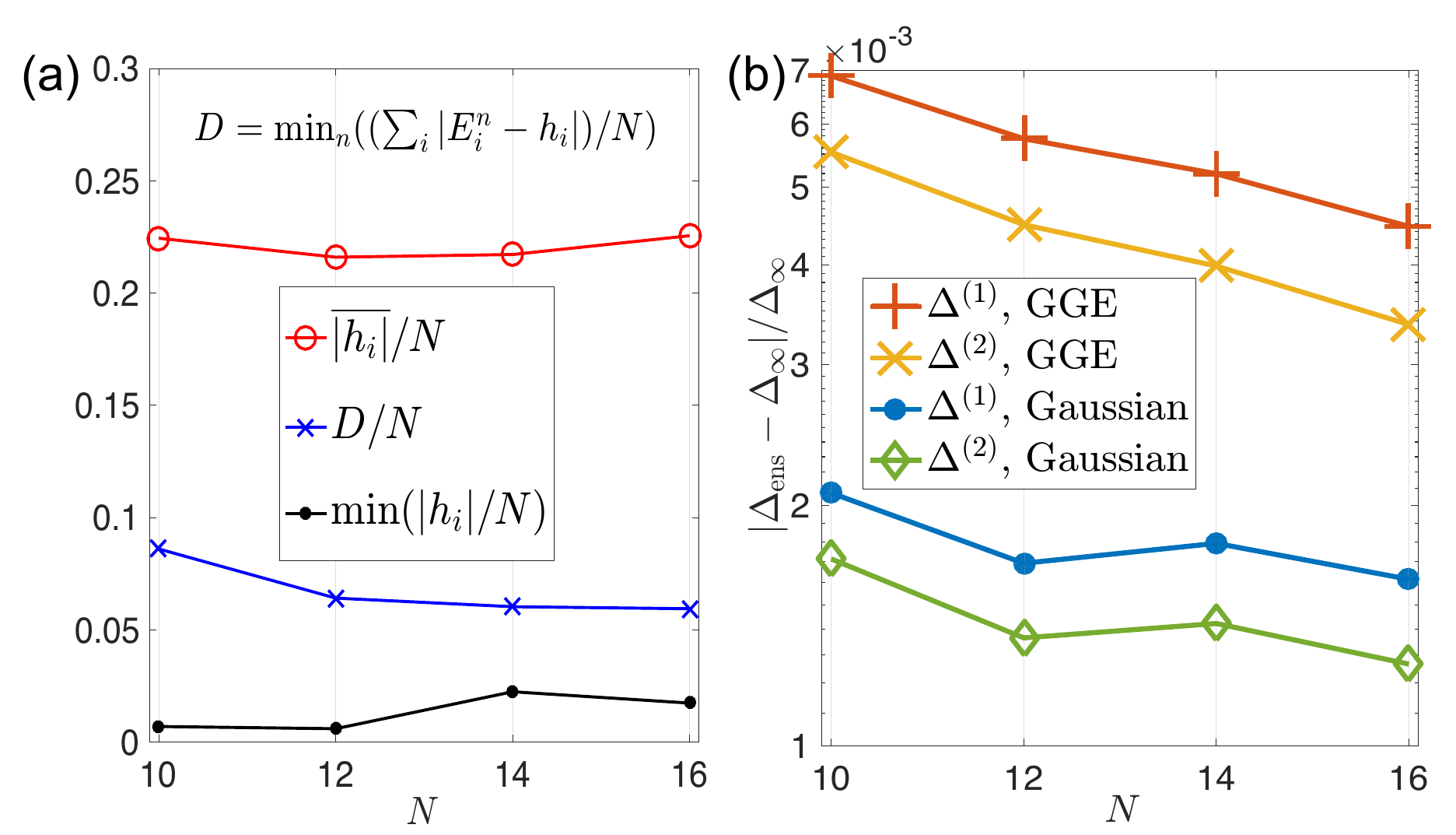}
\centering
\caption{(color online) Interaction quench from $\lambda_\mathrm{in}=0.5$ to $\lambda_\mathrm{fn}=2.0$ in the BCS model \re{bcs} with $N$ single-particle levels (spins). (a) Distance $D$  between the eigenvalues $E_i^{(n)}$ of the integrals of motion and their expectation values $h_i$ in the initial state compared to the arithmetic mean of $|h_i|$ and smallest $|h_i|$. (b) Mismatch between ensemble   (Gaussian and GGE) and infinite time   ($\Delta_\infty$) averages for two alternative definitions, $\Delta^{(1)}$ and  $\Delta^{(2)}$, of the BCS order parameter \re{eq:gaps}.}
\label{fig:BCS}
\end{figure}
{\it Long-range interactions.} 
Our last example is the BCS  model \cite{bcs} on $N$ single-particle levels $\epsilon_i$. In terms of Anderson pseudospin-1/2 operators \cite{anderson},  
\begin{align}
\hat H_\mathrm{BCS} = \sum_{i=1}^N 2\epsilon_i \hat S_i^z - g \sum_{i,j = 1}^N \hat S_i^+ \hat S_j^- .  
\label{bcs}
\end{align}
As usual, we write the BCS coupling constant as $g=\lambda \delta$ \cite{kurland,delft}, 
where $\delta$ is the mean  spacing between $\epsilon_i$'s and $\lambda$ is the dimensionless BCS coupling strength.  

The integrals of motion of the BCS Hamiltonian are  Gaudin magnets \cite{gaudin,integ} (central spin models),
\begin{align}
\hat H_i =-\frac{1}{g}\hat S_i^z + \sum_{j \neq i}^{N} \frac{\hat {\bm S}_i\cdot\hat{\bm S}_j}{\epsilon_i - \epsilon_j},\quad  [\hat H_i, \hat H_j]=0,
\label{gaudin}
\end{align}
It is straightforward to verify that  the total $z$-projection $\hat S^z = -g \sum_i \hat H_i$  and  $\hat H_\mathrm{BCS} = \sum_{i=1}^N 2\epsilon_i H_i + \textrm{const}$. Therefore, $[\hat H_\mathrm{BCS},\hat H_i] =0$ and $\hat S^z$ is conserved. 

This model has several interesting features in addition to being the celebrated BCS model of superconductivity. First, $\hat H_i$ are conserved for any (even unequal) spin magnitudes $S_i$, not just $S_i=1/2$, i.e. the model is integrable (whatever this means in the quantum case) for arbitrary $S_i$. It is therefore one of the few models where one can gradually go from extreme quantum to purely classical while maintaining integrability. In the classical limit, when spins become angular momenta variables and commutators turn into Poisson brackets, $H_\mathrm{BCS}$ is integrable in the strict classical sense \cite{us1,us} and \eref{clGME2} holds. Thus, the Gaussian ensemble \re{eq:GME} has a good foundation. On the other hand, the range of interactions in \eref{bcs} is infinite and $\hat H_i$ are not additive. So, there is no  obvious justification \cite{landau} for a factorizable  exponential form of the density matrix in \eref{gge}.

 We  perform a sudden interaction quench $\lambda_\mathrm{in} \to\lambda_\mathrm{fn}$, i.e. we evolve with $\hat H_\mathrm{BCS}(\lambda_\mathrm{fn})$ starting from the ground state at $\lambda_\mathrm{in}$. In numerics, we take $N=2K$,  $S^z = 0$, $\lambda_\mathrm{in} = 0.5$, and $\lambda_\mathrm{fn} = 2.0$.  The single-particle spectrum is $\epsilon_i = 2i/(N-1)+\eta_i$, where $\eta_K=\eta_{K+1}=0.1$ and $\eta_i=0$ for $i\ne K, K+1$ \cite{asymmetry}.
Precise choice of  the parameters is unimportant 
as long as the  initial  state is not dominated by just a few eigenstates of $\hat H_\mathrm{BCS}(\lambda_\mathrm{fn})$.
 Otherwise, Gaussian ensemble becomes effectively exact as it has enough parameters to match   all  $|c_n|^2$ in time averages $\overline{\langle\hat O\rangle}_\infty=\sum_n |c_n|^2 \hat O_{nn}$, where $c_n$ are the coefficients in the decomposition of the initial state into the eigenstates of $\hat H_\mathrm{BCS}(\lambda_\mathrm{fn})$.
This is not a problem for large $N$, but for $N \leq 12$ we need to choose $\lambda_\mathrm{in}$ and $\lambda_\mathrm{fn}$ carefully to satisfy this condition. 
We use  exact numerical diagonalization to construct the Gaussian ensemble, GGE, and determine $|c_n|^2$ for $N \leq 16$. 
The main challenge turns out to be solving for the Gaussian ensemble and GGE parameters, not  obtaining the eigenstates.  

Our attempts to construct an equal weight  ensemble failed as we were unable to find even a single eigenstates with eigenvalues $E_i^{(n)}$ of $\hat H_i$ sufficiently close to their expectation values $h_i=\langle \hat H_i\rangle_0$ in the initial state. For example, in Fig.~\ref{fig:BCS}(a) we show the minimum distance between $E_i^{(n)}$ and $h_i$,
\beg
D=\frac{1}{N}\min_n\biggl( \sum_{i=1}^N \left| E_i^{(n)}-h_i\right|\biggr),
\en
where $N^{-1}$ ensures $D\propto N$ for large $N$, since $E_i^{(n)}$ and $h_i$ grow as $N$. We see that $D$ is comparable to the average $\overline{h_i}=N^{-1}\sum_i h_i$ and by far exceeds the smallest $|h_i|$. Moreover, $D/N$ does not appreciably decrease with increasing $N$ between $N=12$ and 16, even though available $N$ are too small for a reliable conclusion.

Now let us compare the Gaussian ensemble and GGE to the  quench dynamics. We look at two  versions of the BCS order parameter suitable for a system
 with fixed particle number (fixed $S^z$) \cite{delft}, 
 \beg
\begin{array}{ll}
\displaystyle\Delta^{(1)} &= \displaystyle g \sqrt{\smash[b]{\sum_{i,j} \langle S^+_i S^-_j\rangle - N_\uparrow}\vphantom{\frac{1}{4}} } \\
\\
\displaystyle \Delta^{(2)} &= \displaystyle g \sum_{i} \sqrt{\frac{1}{4} - \langle S^z_{i}\rangle^2}, 
\end{array}
\label{eq:gaps}
\en
where $N_\uparrow$ is the number of up spins, $N_\uparrow = N/2$ for $S^z = 0$. 
Fig.~\ref{fig:BCS}(b) shows  the normalized difference between ensemble averaged $\Delta^{(i=1,2)}$ for both ensembles and the infinite time average [performed before taking square roots in \eref{eq:gaps}] as a function of   $N$. The difference is of order $10^{-2}-10^{-3}$ with an overall decreasing trend with increasing $N$ for either ensemble for both definitions of the order parameter. The mismatch is significantly smaller for the Gaussian ensemble for all $N$.

It is well-known that mean-field becomes exact for the BCS model in $N\to\infty$ limit due to infinite interaction range \cite{anderson,richardson,us2}. Mean-field is equivalent to replacing quantum spin-1/2s with classical spins \cite{anderson,us1,us} in the Hamiltonian \re{bcs} and integrals of motion \re{gaudin}.  On the other hand, we know that  Gaussian ensemble \re{eq:GME} is exact in the classical limit. Thus, we expect it to become exact for the BCS dynamics in the thermodynamic limit. The parameter controlling finite size corrections to  mean-field   is $\delta/\Delta_0$, where $\Delta_0$ is the ground state gap. This parameter therefore plays the role of $\hbar$. If the Gaussian ensemble captures the leading quantum correction of order $\hbar$ to the classical limit as in previous examples,   the discrepancy between ensemble and infinite time averages should go as $(\delta/\Delta_0)^2\propto N^{-2}$. 

 In conclusion, we proposed the multivariable Gaussian ensemble \re{eq:GME} as a quantum extension of exact classical GME  \re{clGME1}. Our proposal stems from the classical definition of integrability thus largely bypassing difficulties associated with the absence of a sound widely accepted quantum notion. It is well and uniformly defined for any quantum system. It is
 exact in the classical limit and provides corrections of order $\hbar$ to this limit at any particle number as well as finite size corrections to GGE  whenever GGE holds. Further, we  expect the Gaussian ensemble to become exact for models with long-range interactions in  thermodynamic limit (as long as mean-field becomes exact). In this paper, we also analyzed two simple one- and two-body models and two many-body models to support these statements.

We thank M. Rigol for  fruitful discussions, especially about GGE. 
H.K. acknowledges technical help from T. Oh. This work was financially supported in part
 by NSF DMR-1308141 (H.K.);  AFOSR FA9550-13-1-0039, NSF DMR-1506340, and ARO W911NF1410540 (A.P.); and David and Lucile Packard Foundation (E.A.Y.).


\begin{thebibliography}{11}
\bibitem{asymmetry}  We make   the spectrum asymmetric with respect to the Fermi energy to avoid degeneracies in the $S^z = 0$ sector due to  particle-hole symmetry.
\end{thebibliography}
\begin{thebibliography}{99}
\bibitem{product} Note that  $\hat \rho=\hat\rho_1\otimes \hat\rho_2$ and $\hat O=\hat O_1\otimes\hat O_2$ implies $\mathrm{tr} (\hat\rho\hat O)=\mathrm{tr} (\hat\rho_1\hat O_1)\mathrm{tr} (\hat\rho_2\hat O_2)$.
\end{thebibliography}

\newpage
\widetext
\begin{center}
\textbf{\large Supplementary Material: Gaussian ensemble for quantum integrable dynamics}
\end{center}

\renewcommand{\theequation}{S\arabic{equation}}
\renewcommand{\thefigure}{S\arabic{figure}}
\renewcommand{\bibnumfmt}[1]{[S#1]}
\renewcommand{\citenumfont}[1]{S#1}

\section{Convergence to the classical limit for 1D harmonic oscillator}

Here we   show that the Gaussian ensemble converges to the classical microcanonical ensemble in $\hbar\to0$ limit and reproduces the leading quantum correction to it for  1D harmonic oscillator in a coherent state.
In this example, there is only one integral of motion (the Hamiltonian itself) in \esref{clGME1} and \re{eq:GME}.  
Suppose the initial state is a coherent state with eigenvalue
$z$: $\hat{a} |z\rangle = z|z\rangle$, where $\hat{a}$ is the annihilation operator.
This corresponds to taking a particle in the ground state of $\hat H_0=(\hat{p} - p_0)^2/(2m)+m \omega^2 (\hat{q} - q_0)^2 / 2 $
and time-evolving with $\hat H= \hat{p}^2/(2m)+m \omega^2 \hat{q}^2 / 2$, where 
$$
z = \sqrt{\frac{m\omega}{2\hbar}}q_0 + i \frac{p_0}{\sqrt{2m\hbar\omega}},
$$
 i.e. to a quantum quench from $p_0, q_0\ne 0$ to $p_0, q_0= 0$
 
To construct $\rho_\mathrm{G}$, we need to fix two parameters, $\mu$ and $\Sigma^{-1}\equiv 2\sigma^2$, with the help of \esref{eq:charge} and \re{eq:correlation} for $\langle \hat H\rangle_0$ and $\langle \hat H^2\rangle_0$. Decomposing $|z\rangle$ into number operator eigenstates $|n\rangle$, we obtain 
\begin{align}
C\sum_{n=0}^\infty \hbar \omega n e^{-\frac{(\hbar\omega n - \mu)^2}{2\sigma^2}} &= \hbar \omega |z|^2 \label{eq:HO_first},\quad (\hat\rho_\mathrm{G})_{nn}=Ce^{-\frac{(\hbar\omega n - \mu)^2}{2\sigma^2}},\\
C\sum_{n=0}^\infty (\hbar \omega n)^2 e^{-\frac{(\hbar\omega n - \mu)^2}{2\sigma^2}} &= (\hbar \omega |z|^2)^2(|z|^2+1),
\label{eq:HO_second}
\end{align}
where $ C^{-1} = \sum_{n=0}^\infty \exp(-(\hbar\omega n - \mu)^2/(2\sigma^2))$ 
    and we include the zero point energy $\hbar\omega/2$ into $\mu$. 
The classical limit is 
\begin{equation}
\hbar\to 0,\quad |z|^2\to \infty,\quad  \hbar \omega|z|^2=\frac{p_0^2}{2m}+\frac{m\omega^2 q_0^2}{2}\equiv E_0=\mbox{fixed}.
\label{cllim}
\end{equation}
In this limit, sums  in \esref{eq:HO_first} and  \re{eq:HO_second} turn into  integrals resulting in $\mu = E_0$ and $\sigma = \hbar \omega |z|$.
 We see that $\sigma\to0$, $\hbar\omega n\to E$ and $\hat \rho_\mathrm{G}\to C\delta(E-E_0)$, i.e. we recover the classical microcanonical ensemble.  

Now let us demonstrate that the Gaussian ensemble captures the leading quantum correction to infinite time averages. We restrict ourselves to powers of the number operator,
$\hat{n}^k=(\hat{a}^\dag\hat{a})^k $, where $k$ is a nonnegative integer. Note that expansion in $\hbar$ is equivalent to the expansion in $1/|z|^2$, see \eref{cllim}. The infinite time average is
\begin{align}
\overline{\langle \hat{n}^k \rangle}_{\infty} =\sum_{n=0}^\infty n^k \left|\langle z|n\rangle\right|^2= |z|^{2k} \left[ 1 + \frac{k(k-1)}{2|z|^2} + O\left(\frac{1}{|z|^4}\right) \right]=|z|^{2k} \left[ 1 + \frac{k(k-1)}{2}\frac{\hbar\omega}{E_0} + O\left(\frac{\hbar^2\omega^2}{E_0^2} \right) \right], 
\label{TAnk}
\end{align}
where we used the fact that $ \left|\langle z|n\rangle\right|^2=|z|^{2n} e^{-|z|^2}/n!$ is a Poisson distribution with parameter $|z|^2$ whose $k^\mathrm{th}$ factorial moment, i.e. the expectation value of $\hat n(\hat n-1)\dots (\hat n-k+1)$, is $|z|^{2k}$, which is straightforward to verify directly. 

Next, we evaluate the Gaussian ensemble averages. Let $n_0 = \mu/(\hbar \omega)$ and $s =  \sigma/(\hbar \omega)$.  We begin by showing that corrections to the classical answer $n_0=s^2=|z|^2$  as  $|z|^2 \rightarrow \infty $ are exponentially small ($\propto e^{-|z|^2}=e^{-E_0/\hbar\omega}$) and therefore can be neglected when calculating corrections of order $\hbar$. 
  \eref{eq:HO_first} becomes
\begin{align}
\frac{\sum_{n=0}^{\infty} n \exp\left(-\frac{(n - n_0)^2}{2s^2}\right)}{\sum_{n=0}^{\infty}  \exp\left(-\frac{(n - n_0)^2}{2s^2}\right)} = |z|^2.
\label{eq:first_moment} 
\end{align}
Observe the following relation for any nonnegative integer $k$ as $n_0, s$, and $n_0/s$ tend to infinity,  
\begin{align}
&\bigg|\sum_{n=0}^{\infty} n^k \exp\left(-\frac{(n - n_0)^2}{2s^2}\right) - \sum_{n=-\infty}^{\infty} n^k \exp\left(-\frac{(n - n_0)^2}{2s^2}\right) \bigg|<  \sum_{n = 2n_0+1}^\infty n^k \exp\left(-\frac{(n - n_0)^2}{2s^2}\right)<\nonumber\\
& \int_{2n_0}^\infty x^k \exp\left(-\frac{(x - n_0)^2}{2s^2}\right) dx =  O\left( s^2 n_0^{k-1} e^{-n_0^2/s^2}\right), 
\end{align}
where we obtained the last relation  by integrating by parts. 
Therefore, we can extend  summations in \eref{eq:first_moment} from $[0,\infty)$ to $(-\infty, \infty)$ with an exponentially small error, i.e. 
\begin{align}
|z|^2 &= \frac{\sum_{n=0}^{\infty} n \exp\left(-\frac{(n - n_0)^2}{2s^2}\right)}{\sum_{n=0}^{\infty}  \exp\left(-\frac{(n - n_0)^2}{2s^2}\right)} = \frac{\sum_{n=-\infty}^{\infty} (n+n_0) \exp\left(-\frac{n^2}{2s^2}\right)}{\sum_{n=-\infty}^{\infty}  \exp\left(-\frac{n^2}{2s^2}\right)}  + O\left(|z|^2e^{-|z|^2}\right) 
= n_0 +  O\left(|z|^2e^{-|z|^2}\right). 
\end{align}
Similarly, we derive $s^2 = |z|^2$ up to an exponentially small error.

Therefore, the Gaussian ensemble expectation value is
\begin{align}
\langle \hat{n}^k \rangle_\mathrm{G} = \frac{\sum_{n=0}^{\infty} n^k \exp\left[-(n - |z|^2)^2/(2|z|^2)\right]}{\sum_{n=0}^{\infty} \exp\left[-(n - |z|^2)^2/(2|z|^2)\right]}.
\end{align}
We evaluate the sums involved using the Poisson summation formula,
\beg
A_k\equiv\sum_{n=0}^{\infty} n^k \exp\left[\frac{-(n - |z|^2)^2}{2|z|^2}\right]=|z|^{2k+2}\sum_{p=-\infty}^\infty
\int_0^\infty e^{-|z|^2\left[ (x-1)^2/2-2i\pi p x\right]} x^k dx.
\en
The saddle-point analysis of the integrals on the r.h.s. shows that the contribution of $p\ne 0$ terms is suppressed by a factor $e^{-2\pi^2 |z|^2}$, i.e. it is sufficient to keep only the $p=0$ term,
\beg
A_k=|z|^{2k+2}\biggl[\int_{-1}^\infty (1+y)^k e^{-|z|^2 y^2/2} dy+ O\left(e^{-2\pi^2 |z|^2}\right)\biggr]
=|z|^{2k}\sqrt{2\pi}|z|\left[1+\frac{k(k-1)}{2}\frac{1}{|z|^2}+O\left(\frac{1}{|z|^4}\right)\right],
\label{akint}
\en
where we changed the variable $x=y+1$ in the integral and evaluated it with a simple version of the saddle-point method. Thus,
\beg
\langle \hat{n}^k \rangle_\mathrm{G}=\frac{A_k}{A_0}=|z|^{2k}\left[ 1 + \frac{k(k-1)}{2|z|^2} + O\left(\frac{1}{|z|^4}\right) \right]=|z|^{2k} \left[ 1 + \frac{k(k-1)}{2}\frac{\hbar\omega}{E_0} + O\left(\frac{\hbar^2\omega^2}{E_0^2} \right) \right].
\label{gaussnk}
\en
Comparing \esref{TAnk} and \re{gaussnk}, we see that they agree up to terms proportional to $\hbar$. In other words,   Gaussian ensemble  reproduces the leading term and the first quantum correction. 

This agreement, however, does not extend to terms of order $\hbar^2$ and higher.  Consider, for example,  $\hat{n}^3$. 
The exact infinite time average  follows from the first three factorial moments mentioned above
\beg
\overline{\langle \hat{n}^3 \rangle}_\infty = |z|^6 + 3|z|^4 + |z|^2=|z|^6\left[1+3\frac{\hbar\omega}{E_0}+ \frac{\hbar^2\omega^2}{E^2_0}\right].
\en
To obtain the Gaussian ensemble, we evaluate the integral in the first equation in \re{akint} for $k=3$ and $k=0$,
\beg
\langle \hat{n}^3 \rangle_\mathrm{G} = |z|^6 + 3|z|^4 +  O\bigl(e^{-|z|^2/2}\bigr)=|z|^6\left[1+3\frac{\hbar\omega}{E_0}+  O\bigl(e^{-E_0/\hbar\omega}\bigr)\right],
\en
where the error arises from $|z|^{2k}\int_{-\infty}^{-1} (1+y)^k e^{-|z|^2 y^2/2}=O\bigl(e^{-|z|^2/2}\bigr)$, which we estimated via repeated integration by parts.
Therefore,  there is a discrepancy of order $\hbar^2$.  For instance,
since $\langle \hat{q}^6 \rangle = (5/2) (\hbar/m\omega)^3 \langle \hat{n}^3\rangle +$  expectation values of lower powers of $\hat{n}$ that 
are captured exactly by construction, 
the discrepancy in the third order cumulant of $\hat{q}^2$ between the Gaussian ensemble and exact infinite time averages is  $2.5(\hbar/m\omega)^3|z|^2=2.5\hbar^2 E_0/m^3\omega^4$.

\section{Calculation of three-body observables in the free-fermion model}

In the main text, we compared GGE and the Gaussian ensemble for quenches in a free-fermion model
\begin{align}
\hat H = &-\sum_{j=1}^L (e^{i \phi} \hat c_{j+1}^\dag \hat c_j +e^{-i\phi} \hat c_j^\dag \hat c_{j+1}) + \sum_{j=1}^L[V_1 \cos(Q j) + V_2 \cos(2Q j)] \hat n_j, 
\end{align}
 with periodic boundary conditions and $Q = 2\pi/M$ with   integer $M$, such that  $L$ is a multiple of $M$. 
The pre-quench Hamiltonian $\hat H_\mathrm{in}$ has $\phi = V_1 = V_2 = 0$ and  is therefore diagonal in the momentum basis,
\begin{align}
\hat H_\mathrm{in} = - \sum_{k} 2\cos(k) \hat \tc_k^\dag \hat \tc_k, 
\end{align}  
where $\hat \tc_k =  L^{-1/2} \sum_{j = 1}^L \hat c_j e^{-i k j}$ and $k = 2\pi \nu/L$ with $\nu =  1,2,  \ldots, L$. 

Due to our choice of the  modulation wavenumber $Q = 2\pi/M$, the quenched Hamiltonian $\hat H_\mathrm{fn}$  mixes momenta $k, k+Q, k+2Q, \ldots k + (M-1)Q$ only among themselves.   It is convenient to introduce  a two-index momentum notation that reflects this property: 
$\hat \tc_k \rightarrow \hat \tc_{q,\alpha}$, where $q = 2\pi/L, 4\pi/L, \ldots, 2\pi/M$,  $\alpha = 0, 1, \ldots, M-1$, and $k = q + \alpha Q$. 
The quenched Hamiltonian  splits into $L/M$ independent sub-Hamiltonians (sectors)
\begin{align}
\hat H_\mathrm{fn} = \sum_{q} \sum_{\alpha,\beta=0}^{M-1} h^q_{\alpha,\beta}\hat\tc_{q,\alpha}^\dag \hat \tc_{q,\beta} = \sum_{q} \sum_{\gamma=0}^{M-1} \eps(q,\gamma) \hat N_{q,\gamma},\quad \hat N_{q,\gamma}=\hat b^\dag_{q,\gamma} \hat b_{q,\gamma}, 
\label{hfn}
\end{align}
where $h^q_{\alpha,\beta}$ is an $M \times M$ matrix,
\begin{align}
{\bm h}^q = 
\begin{pmatrix}
-2 \cos(q+\phi) && V_1/2 && V_2/2 &&\cdots && V_1/2 \\
V_1/2 && -2\cos(q+\phi + Q) && V_1/2 && \cdots && V_2/2\\
V_2/2 && V_1/2 && -2\cos(q + +\phi + 2Q) && \cdots &&\vdots  \\
\vdots && \vdots && \vdots && \ddots && V_1/2 \\
V_1/2 && \cdots && V_2/2 && V_1/2 && -2\cos(q + \phi+ (M-1)Q) 
\end{pmatrix}.
\end{align}
 Diagonalizing  ${\bm h}^q$, we obtain  single-particle energies ${\bm \epsilon}(q) = ({\bf U}^q)^{-1} {\bf h}^q {\bf U}^q$ and  new fermion operators  $\hat b_{q,\gamma} = \sum_\beta (U^q)^{-1}_{\gamma,\beta} \hat \tc_{q,\beta}$.  
 
 The conservation laws are mode occupation numbers $\hat N_{q,\gamma}$. The GGE and Gaussian ensemble density matrices are
 \begin{align}
 \hat\rho_\mathrm{GGE}=\exp\biggl[-\sum_{k,\alpha} \lambda_{k,\alpha} \hat N_{k,\alpha}   \biggr],\quad \langle\hat N_{k,\alpha}\rangle_\mathrm{GGE}\equiv\frac{\mathrm{tr}\left(\hat \rho_\mathrm{GGE} \hat N_{k,\alpha}\right)}{\mathrm{tr }\,\hat\rho_\mathrm{GGE}}=\langle\hat N_{k,\alpha}\rangle_0,\label{matr}\\ 
  \hat\rho_\mathrm{G}= \exp\biggl[-\!\!\!\sum_{p,q,\alpha,\beta} \sigma_{p\alpha,q\beta} \hat N_{p,\alpha}    \hat N_{q,\beta} \biggr],\quad \langle\hat N_{p,\alpha}\rangle_\mathrm{G}\equiv\frac{\mathrm{tr}\left(\hat \rho_\mathrm{G} \hat N_{p,\alpha}\right)}{\mathrm{tr }\,\hat\rho_\mathrm{G}}=\langle\hat N_{p,\alpha}\rangle_0,\quad 
  \langle\hat N_{p,\alpha}    \hat N_{q,\beta}\rangle_\mathrm{G}=\langle\hat N_{p,\alpha}    \hat N_{q,\beta}\rangle_0,\label{Gmatr}
 \end{align}
where we used $ \hat N_{q,\beta}= \hat N_{q,\beta}^2$ to absorb the linear in $\hat N_{q,\beta}$ terms in  the definition of $\hat\rho_\mathrm{G}$ into the quadratic part.
 
We are interested in the three-point   correlation function of the lattice site occupation number
\begin{align}
\langle \hat n_j \hat n_\ell \hat n_m \rangle&= \langle \hat c^\dag_j \hat c_j \hat c^\dag_\ell \hat c_\ell \hat c_m^\dag \hat c_m \rangle \\
& =  \frac{1}{L^3} \sum_{k,q,u,v,p,s}\sum_{\alpha,\beta,\gamma,\delta,\zeta,\eta}e^{i[k - q + (\beta - \alpha)Q]j + i[u - v + (\delta - \gamma)Q]\ell + i[s - p + (\eta - \zeta)Q]m}\langle \hat\tc^\dag_{q,\alpha} \hat\tc_{k,\beta} \hat\tc^\dag_{v,\gamma} \hat\tc_{u,\delta} \hat\tc^\dag_{p,\zeta} \hat\tc_{s,\eta}\rangle\\
&=  \frac{1}{L^3} \sum_{k,q,u,v,p,s}\sum_{\alpha,\beta,\gamma,\delta,\zeta,\eta}e^{i[k - q + (\beta - \alpha)Q]j + i[u - v + (\delta - \gamma)Q]\ell + i[s - p + (\eta - \zeta)Q]m}  \times \nonumber\\
&\quad\quad\sum_{\alpha',\beta',\gamma',\delta',\zeta',\eta'} (U^{q})^*_{\alpha,\alpha'}  (U^k)_{\beta,\beta'}  (U^{v})^*_{\gamma,\gamma'} (U^{u})_{\delta,\delta'}(U^p)^*_{\zeta,\zeta'} (U^s)_{\eta,\eta'}\langle \hat b^\dag_{q,\alpha'}\hat b_{k,\beta'} \hat b^\dag_{v, \gamma'} \hat b_{u,\delta'} \hat b^\dag_{p,\zeta'} \hat b_{s,\eta'}\rangle. 
\label{eq:full_three_body}
\end{align}
Therefore, we need to evaluate the following average: 
\begin{align}
\langle \hat b^\dag_{q,\alpha'} \hat b_{k,\beta'} \hat b^\dag_{u,\gamma'} \hat b_{v,\delta'} \hat b^\dag_{p,\zeta'}\hat b_{s,\eta'} \rangle, 
\label{eq:three_body}
\end{align} 
with respect to the time-evolved state of the system as well as Gaussian ensemble and GGE.

The time-evolution  is
\begin{align}
\langle \hat b^\dag_{q,\alpha'}\hat  b_{k,\beta'} \hat b^\dag_{v, \gamma'} \hat b_{u,\delta'} \hat b^\dag_{p,\zeta'} \hat b_{s,\eta'} \rangle_t  =  e^{i[\epsilon(q,\alpha') - \epsilon(k,\beta') + \epsilon(v,\gamma') - \epsilon(u,\delta')) + \epsilon(p,\zeta') - \epsilon(s,\eta')]t} \langle \hat b^\dag_{q,\alpha'}\hat b_{k,\beta'} \hat b^\dag_{v, \gamma'} \hat b_{u,\delta'} \hat b^\dag_{p,\zeta'} \hat b_{s,\eta'} \rangle_0, 
\end{align}
where $\langle \ldots \rangle_t$ is the expectation value at time $t$. 
The infinite time average  is nonzero only for terms with zero phase factor, i.e. when $ \epsilon(q,\alpha') - \epsilon(k,\beta') + \epsilon(v,\gamma') - \epsilon(u,\delta') + \epsilon(p,\zeta') - \epsilon(s,\eta') = 0$. Since we  removed time-reversal   and particle-hole symmetries, there  are no
degeneracies in the single-particle spectrum as well as no two-particle and three-particle resonances. 
Consequently, the time average  is zero unless the double indices on creation and annihilation operators are pairwise equal, i.e. the set $\{(q,\alpha'), (v,\gamma'), (p,\zeta')\}$ is a permutation of $\{(k,\beta'),  (u,\delta'), (s,\eta')\}$. Therefore, only terms that can be cast into the form $\langle \hat N_{q,\alpha}\hat  N_{r,\beta} \hat N_{s, \gamma}   \rangle_t=\langle \hat N_{q,\alpha}\hat  N_{r,\beta} \hat N_{s, \gamma}   \rangle_0$ survive. The same holds for expectation value \re{eq:three_body} evaluated in any eigenstate of $\hat H_\mathrm{fn}$ and hence for both ensemble averages.  Thus,
\beg
\overline{\langle \hat n_j \hat n_\ell \hat n_m \rangle}_\infty-\langle \hat n_j \hat n_\ell \hat n_m \rangle_\mathrm{ens}=\frac{1}{L^3}\!\!\!\sum_{q,r,s;\alpha,\beta,\gamma} R^{qrs}_{\alpha\beta\gamma}
\left(\langle \hat N_{q,\alpha}\hat  N_{r,\beta} \hat N_{s, \gamma}   \rangle_0-\langle \hat N_{q,\alpha}\hat  N_{r,\beta} \hat N_{s, \gamma}   \rangle_\mathrm{ens}\right),
\label{diff}
\en
where $R^{qrs}_{\alpha\beta\gamma}$ are coefficients of order one and $\langle\dots\rangle_\mathrm{ens}$ stands for the average with respect to either ensemble.

Because $\hat \rho_\mathrm{GGE}$ in \eref{matr} is a (tensor) product of functions of individual occupation numbers \cite{product},
\beg
\langle \hat N_{q,\alpha}\hat  N_{r,\beta} \hat N_{s, \gamma}   \rangle_\mathrm{GGE}=
\langle \hat N_{q,\alpha}\rangle_\mathrm{GGE}\langle\hat  N_{r,\beta} \rangle_\mathrm{GGE}\langle\hat N_{s, \gamma}   \rangle_\mathrm{GGE}=\langle \hat N_{q,\alpha}\rangle_0\langle\hat  N_{r,\beta} \rangle_0\langle\hat N_{s, \gamma}   \rangle_0,
\label{GGEfactor}
\en
as long as no two occupation number operators, i.e.  \textit{pairs} of indices coincide,  $\{q,\alpha\}\ne \{r,\beta\}\ne \{s, \gamma\}$ and $\{q,\alpha\} \ne \{s, \gamma\}$. Since occupation numbers in different sectors are uncorrelated in the initial state, the same factorization holds for $\langle \hat N_{q,\alpha}\hat  N_{r,\beta} \hat N_{s, \gamma}   \rangle_0$ if $q, r, s$ are distinct. For most remaining terms,  when $q=r$ (but $\alpha\ne\beta$) or $q=s$ (but $\alpha\ne\gamma$) or $r=s$ (but $\beta\ne\gamma$), the GGE and initial state averages do not necessarily agree. In other words,  GGE fails to capture the correlations between different occupation numbers within the same sector. Since the number of such terms($\propto$ number of  pairs of sectors) is of order $(L/M)^2\propto L^2$, $\overline{\langle \hat n_j \hat n_\ell \hat n_m \rangle}_\infty-\langle \hat n_j \hat n_\ell \hat n_m \rangle_\mathrm{GGE}\propto 1/L$ at large $L$ ($M$ is fixed). Similarly, the Gaussian ensemble  automatically matches $\langle \hat N_{q,\alpha}\hat  N_{r,\beta} \hat N_{s, \gamma}   \rangle_0$ for distinct $q, r, s$ and for
$q=r\ne s$, $q\ne r=s$,  $q=s\ne r$, but not for $q=r=s$ (except when two of the indices $\alpha,\beta,\gamma$ are equal).  In other words, it reproduces two-body, but not three-body correlations between occupation numbers in the initial state. The number of $q=r=s$ terms is proportional to $L$, so  $\overline{\langle \hat n_j \hat n_\ell \hat n_m \rangle}_\infty-\langle \hat n_j \hat n_\ell \hat n_m \rangle_\mathrm{G}\propto 1/L^2$. This behavior with the system size $L$ is in agreement with Fig.~3 in the main text for both ensembles. 
The fact that the Gaussian ensemble is exact for  one- and two-point correlation functions of lattice site occupation numbers forced us to consider  three-point functions.  

\subsection{Computing initial state and ensemble averages}

Another implication of \eref{GGEfactor} is that we need not determine the Lagrange multiplies $\lambda_{q,\alpha}$ for GGE in \eref{matr}, since GGE averages  reduce to expectation values of single mode occupation numbers $\langle\hat N_{k,\alpha}\rangle_0$  in the initial state. 

We do however need  $\langle\hat N_{k,\alpha}\rangle_0$ and we also need 
 $\langle\hat N_{p,\alpha}    \hat N_{q,\beta}\rangle_0$ in \esref{Gmatr} to construct the Gaussian ensemble,
\begin{align}
  \langle\hat N_{k,\alpha}\rangle_0= \langle 0| \prod_q \hat\tc_q (\hat b_{k,\alpha}^\dag \hat b_{k,\alpha}) \prod_{p} \hat\tc^\dag_p|0\rangle &=  \sum_\beta |(U^k)^{-1}_{\alpha,\beta}|^2\langle 0| \prod_q \hat \tc_q (\hat \tc^\dag_{k,\beta} \hat \tc_{k,\beta})  \prod_p \hat\tc^\dag_p|0\rangle 
= \!\!\!\sum_{\beta: k + \beta Q \in \{\mathrm{gr}\}}\!\!\! |(U^k)^{-1}_{\alpha,\beta}|^2, 
\end{align}
\begin{align}
 \langle\hat N_{p,\alpha}    \hat N_{q,\beta}\rangle_0 &= \langle 0| \prod_u \hat\tc_u (\hat b_{p,\alpha}^\dag \hat b_{p,\alpha} \hat b_{q,\beta}^\dag \hat b_{q,\beta}) \prod_v \hat\tc^\dag_v|0\rangle =\nonumber\\
& \sum_{\gamma,\gamma'\!,\delta,\delta'} (U^p)^{-1}_{\alpha,\gamma} (U^{p*})^{-1}_{\alpha,\gamma'} (U^q)^{-1}_{\beta,\delta} (U^{q*})^{-1}_{\beta,\delta'} \langle 0| \prod_u \hat \tc_u (\hat\tc^\dag_{p,\gamma'} \hat\tc_{p,\gamma} \hat\tc^\dag_{q,\delta'} \hat\tc_{q,\delta}) \prod_v \hat\tc^\dag_v|0\rangle=\nonumber\\
&\sum_{q + \delta Q, p + \gamma Q \in \{\mathrm{gr}\}} |(U^p)^{-1}_{\alpha,\gamma}|^2 |(U^q)^{-1}_{\beta,\delta}|^2  + \sum_{\substack{q + \delta Q\in \{\mathrm{gr} \} \\ q + \gamma Q\not\in \{\mathrm{gr} \}}} (U^q)^{-1}_{\alpha,\gamma} (U^{q*})^{-1}_{\beta,\gamma} (U^q)^{-1}_{\alpha,\delta} (U^{q*})^{-1}_{\beta,\delta}, 
\end{align} 
where $\{ \mathrm{gr}\}$ is the set of momenta occupied in the ground state of the pre-quench Hamiltonian. 

To construct the Gaussian ensemble, we have to solve the last two equations in \re{Gmatr} for $ \sigma_{p\alpha,q\beta}$. Since $\langle\hat N_{p,\alpha}    \hat N_{q,\beta}\rangle_0=\langle\hat N_{p,\alpha}\rangle_0\langle    \hat N_{q,\beta}\rangle_0$ for $p\ne q$ in the initial state, we set $ \sigma_{p\alpha,q\beta}=0$ for $p\ne q$. Then, $\hat\rho_\mathrm{G}$ is a tensor product over different sectors labeled by $p$, which ensures \cite{product} $\langle\hat N_{p,\alpha}    \hat N_{q,\beta}\rangle_\mathrm{G}=\langle\hat N_{p,\alpha}\rangle_\mathrm{G}\langle    \hat N_{q,\beta}\rangle_\mathrm{G}$. Thus, we are left with
\beg
 \frac{\mathrm{tr}\left(\hat \rho_\mathrm{G} \hat N_{p,\alpha}\right)}{\mathrm{tr }\,\hat\rho_\mathrm{G}}=\langle\hat N_{p,\alpha}\rangle_0,\quad 
 \frac{\mathrm{tr}\left(\hat \rho_\mathrm{G} \hat N_{p,\alpha}    \hat N_{p,\beta}\right)}{\mathrm{tr }\,\hat\rho_\mathrm{G}}  =\langle\hat N_{p,\alpha}    \hat N_{p,\beta}\rangle_0,\quad  \hat\rho_\mathrm{G}= \exp\biggl[-\!\!\!\sum_{p,\alpha,\beta} \sigma_{p\alpha,p\beta} \hat N_{p,\alpha}    \hat N_{p,\beta} \biggr].
  \label{Gmatr1}
 \en
 There are $M(M+1)/2$ nonlinear equations for  $M(M+1)/2$ unknown $\sigma_{p\alpha,p\beta}$  for each of $L/M$ sectors labeled by $p$, a total of $L(M+1)/2$ equations, to be solved numerically. This is feasible for moderate $M$($\le 12$).  
 
 Note also that the number of particles in each sector [each sub-Hamiltonian in \eref{hfn}] is conserved. Therefore, the number of eigenstates $|n\rangle$ of $H_\mathrm{fn}$ involved in the time-evolution in each sector is
 $C^M_m$. The infinite time average of an observable $\hat O$ in the absence of degeneracies is $\overline{\langle \hat O\rangle}_\infty=\sum_n |c_n|^2\langle n|\hat O|n\rangle$, where $c_n$ are the coefficients in the decomposition of the initial state into the eigenstates (diagonal ensemble). We need  $M(M+1)/2>C^M_m$ or else the Gaussian ensemble becomes exact as it has enough parameters to match all $|c_n|^2$. 
The smallest $M$ that satisfies this criterion is $M=6$ at  $m = 3$. This dictates our choice of $M = 6$ and $\mbox{filling fraction}=1/2$. 

Finally, we comment on a technical aspect of the computation. Even though we are dealing with a 
  free-fermion model, a direct computation of three-point correlation functions (unlike two- and one-point ones) is prohibitive at large $L(\geq 120)$ due to a large number $(\propto L^3)$ of nonzero terms in \eref{eq:full_three_body}. However,  only a small fraction of these terms   contribute to the  difference between ensemble and infinite time averages in \eref{diff}.  As discussed above, there is a factor of $(L/M)^2$ reduction in the number of terms for the Gaussian ensemble and a factor of $L/M$ for GGE. This allows as to go to much larger $L$ in numerical evaluation of the difference.

\end{document}